\documentclass[twocolumn,aps]{revtex4}
\usepackage{amsmath}
\usepackage{amssymb}
\usepackage{graphicx}
\usepackage{dcolumn}
\usepackage{bm}
\usepackage{xcolor}
\usepackage{epstopdf}

\begin{document}

\title{Manifestations of classical size effect and electronic viscosity 
in magnetoresistance of narrow two-dimensional conductors: theory and 
experiment}

\author{O. E. Raichev,$^1$ G. M. Gusev,$^2$ A. D. Levin,$^2$ 
and A. K. Bakarov$^{3,4}$}

\affiliation{$^1$Institute of Semiconductor Physics, NAS of
Ukraine, Prospekt Nauki 41, 03028 Kyiv, Ukraine}
\affiliation{$^2$Instituto de F\'{\i}sica da Universidade de S\~ao
Paulo, 135960-170, S\~ao Paulo, SP, Brazil}
\affiliation{$^3$Institute of Semiconductor Physics, Novosibirsk
630090, Russia}
\affiliation{$^4$Novosibirsk State University, Novosibirsk 630090, Russia}

\date{\today}

\begin{abstract}
We develop a classical kinetic theory of magnetotransport of 2D electrons in narrow 
channels with partly diffusive boundary scattering and apply it to description of magnetoresistance 
measured in the temperature interval 4.2-30 K in long mesoscopic bars fabricated from 
high-purity GaAs quantum well structures. Both experiment and theory demonstrate a number 
of characteristic features in the longitudinal and Hall resistances caused by the size 
effect in two dimensions owing to the high ballisticity of the transport. In addition to the 
features described previously, we also reveal a change in the slope of the first derivative 
of magnetoresistance when the cyclotron orbit diameter equals to half of the channel width. 
These features are suppressed with increasing temperature as a result of the electronic 
viscosity due to electron-electron interaction. By comparing theory and experiment, we 
determine the characteristic time of relaxation of angular distribution of electrons 
caused by electron-electron scattering.  
 
\end{abstract}

\maketitle

\section{Introduction}

In past years, numerous experimental and theoretical studies have revealed 
interesting effects in transport of two-dimensional (2D) electron systems 
under conditions when electron movement is affected by internal friction 
due to interaction between the particles and resembles the dynamics 
of viscous fluids \cite{gurzhi}-\cite{holder}. Such effects become important 
even in the linear transport, provided that the electron system is spatially 
inhomogeneous and electron-electron interaction is sufficiently strong. 
The hydrodynamic transport regime can be detected, in particular, in narrow 
conducting channels (2D wires), when the mean free path of electrons with 
respect to momentum changing scattering by impurities and phonons, $l_1$, is 
larger than the channel width $L$, while the mean free path with 
respect to momentum conserving electron-electron scattering, $l_{e}$, is much 
smaller than both $l_1$ and $L$. Due to the dominance of electron-electron scattering 
over the other scattering processes, the standard Drude picture of transport 
becomes invalid. As it was found in the pioneering theoretical study by Gurzhi 
\cite{gurzhi}, in these conditions the ohmic resistivity should decrease with 
increasing temperature $T$ in a certain interval of $T$ and depend on the 
channel width. In 2D systems, a temperature-induced decrease of resistivity, 
attributed to the Gurzhi effect, was observed under conditions when electrons 
were heated by the current \cite{dejong}, and in a special (H-shaped) bar geometry 
\cite{gusev1}. More manifestations of electron viscosity in narrow 2D channels can 
be found in the presence of a transverse magnetic field $B$. 

The features of narrow channel resistance associated with hydrodynamic transport 
are easier to observe in the systems with a large mean free path $l_1$, such as 
graphene and high-purity GaAs quantum wells with large electron densities $n_s 
\sim 10^{12}$ cm$^{-2}$, though in both cases one requires elevated electron 
temperatures $T \sim 100$ K to enable strong electron-electron scattering. 
At lower temperatures, the transport regime is intermediate 
between hydrodynamic and quasi-ballistic regimes. A purely hydrodynamic approach to 
transport, implying a solution of the linearized Navier-Stokes equation with boundary 
conditions for electron current or drift velocity \cite{alekseev1}, is insufficient 
in this case. Thus, a description of transport properties should be based on a more 
detailed approach assuming solution of the Boltzmann kinetic equation complemented 
with the boundary conditions for the electron distribution function. The kinetic equation 
approach is valid for an arbitrary hierarchy of the characteristic lengths $l_{e}$, 
$l_1$, and $L$, so the standard diffusive (Drude), ballistic (Knudsen), and hydrodynamic 
(Poiseuille) transport regimes follow as limiting cases of the general description. 
With a simplifying relaxation-time approximation for the electron-electron collision 
integral, the kinetic equation is reduced to a differential equation and allows for 
either analytical or numerical solution \cite{dejong}, \cite{govorov}, \cite{scaffidi}, 
\cite{lucas1}, \cite{lucas2}, \cite{alekseev2}, \cite{alekseev3}, \cite{chandra}, 
\cite{holder}. In the presence of a magnetic field, however, the problem still 
remains complicated, as the kinetic equation is a partial differential equation 
involving the derivatives over both spatial coordinates and electron momentum. 
This problem has been recently solved in the geometry of an infinitely 
long 2D channel, when the distribution function depends only on 
one spatial coordinate. A numerical solution has been obtained by using the method of 
characteristics together with the boundary conditions for fully diffusive scattering 
on the edges (boundaries) \cite{scaffidi}. An approximate perturbative solution with 
similar boundary conditions has been found for the case of small magnetic fields 
\cite{alekseev2}, \cite{alekseev3}. A numerical solution by the method of characteristics 
has also been obtained for a more realistic case of partly diffusive scattering at the 
edges \cite{holder}. However, the boundary conditions proposed in Ref. \cite{holder} 
are not justified from a microscopic consideration of electron scattering at the edge 
and do not guarantee the necessary requirement of zero flux of electrons through the edge. 

In this paper, we further develop the theory of magnetotransport in narrow conducting 
channels by applying reliable boundary conditions for solution of the kinetic equation. 
Then we carry out a detailed comparison of the results of theoretical calculations with 
experimental magnetotransport data, which has not been done in previous works. Such 
a comparison allows us to investigate both the classical size effect and the influence of 
viscosity on magnetotransport properties in a wide temperature range on an equal footing. 
This leads us to a deeper understanding of the roles of boundary scattering and 
electron-electron interaction in transport of bounded 2D fermion systems and provides 
an estimate for electron-electron scattering time characterizing momentum relaxation 
of electron distribution.  

The paper is organized as follows. In Sec. II we describe the theoretical model and present 
some results of its application. Section III contains description of measurements, presentation 
of experimental and theoretical results, their comparison and discussion. More discussion and 
concluding remarks are given in the last section. The Appendix provides the 
details of the solution of the kinetic equation by the method of characteristics. 

\section{Theory}

The classical kinetic equation for the distribution function $f_{{\bf p}}({\bf r})$ in 
the electric field ${\bf E}({\bf r})=-\nabla \Phi({\bf r})$ ($\Phi$ is the electrostatic potential) 
and homogeneous magnetic field ${\bf B}$ directed perpendicular to the 2D plane is 
\begin{eqnarray}
{\bf v} \cdot \nabla f_{{\bf p}}({\bf r})
+ \left(e {\bf E}({\bf r}) + 
\frac{e}{c}[{\bf v} \times {\bf B}] \right) \cdot
\frac{\partial}{\partial {\bf p}} f_{{\bf p}}({\bf r}) 
= {\cal J}_{{\bf p}}({\bf r}),   
\end{eqnarray}
where ${\bf r}=(x,y)$ and ${\bf p}$ are the coordinate and momentum of 
electrons, $e$ is the electron charge, and $c$ is the light velocity.
For electrons with isotropic and parabolic spectrum, the velocity is 
given by ${\bf v}={\bf p}/m$, where $m$ is the effective mass. The 
right-hand side of Eq. (1) contains the collision integrals specified below. 
Instead of two components of ${\bf p}$, it is convenient to use energy and 
angle variables according to ${\bf p}= m v_{\varepsilon} (\cos 
\varphi, \sin \varphi)$ so that $f_{{\bf p}}({\bf r}) \equiv 
f_{\varepsilon \varphi}({\bf r})$, where $\varphi$ is the angle 
between the $x$ axis and the direction of momentum.  

Assume that there is a boundary $y=y_0$ and electrons occupy the region above 
the boundary, $y>y_0$. If boundary scattering of electrons is elastic and not 
influenced by the magnetic field, the most general boundary condition for the 
distribution function $f_{\varepsilon \varphi}({\bf r})$ at the boundary 
${\bf r}=(x,y_0)$ takes the form 
\begin{eqnarray}
f_{\varepsilon \varphi}= {\rm r}_{\varepsilon \varphi} f_{\varepsilon 2 \pi-\varphi}  + 
\int_{0}^{\pi} \frac{d \varphi'}{\pi} \sin \varphi' P_{\varepsilon}(\varphi,\varphi') 
f_{\varepsilon 2 \pi-\varphi'}, \\
{\rm r}_{\varepsilon \varphi} = 1 - \int_{0}^{\pi} \frac{d \varphi'}{\pi} \sin \varphi' 
P_{\varepsilon}(\varphi,\varphi') ,~ \varphi \in [0,\pi]. \nonumber
\end{eqnarray}
The left-hand side of this equation presents the distribution function of 
reflected electrons, for which $\varphi \in [0,\pi]$. The right-hand side is 
expressed through the distribution function of incident electrons, part of which 
is reflected specularly. The probability of specular scattering is characterized 
by the reflection coefficient ${\rm r}_{\varepsilon \varphi}$. 
The function $P_{\varepsilon}(\varphi,\varphi')$ is determined by the properties 
of boundary scattering. It is symmetric with respect to permutation of
variables, $P_{\varepsilon}(\varphi,\varphi')=P_{\varepsilon}(\varphi',\varphi)$, 
and goes to zero at $\varphi=0$ and $\varphi=\pi$ because the boundary does not 
affect the electrons moving parallel to it. Equation (2) can be obtained by 
a direct adoption of the boundary conditions derived for three-dimensional 
electrons \cite{soffer}, \cite{falkovski}, \cite{okulov} to the case of 2D 
electrons. This equation automatically guarantees zero particle flux 
through the boundary, $\int^{2 \pi}_0 d \varphi v_y f_{\varepsilon \varphi} 
=v_{\varepsilon} \int^{\pi}_0 d \varphi \sin \varphi(f_{\varepsilon \varphi}-
f_{\varepsilon 2 \pi-\varphi})=0$. Under certain conditions, the symmetry 
of the distribution function makes the integral term in Eq. (2) equal to zero, 
and the boundary condition takes a simple form, $f_{\varepsilon \varphi}= {\rm r}_{
\varepsilon \varphi} f_{\varepsilon 2 \pi-\varphi}$, similar to that proposed 
by Fuchs \cite{fuchs}. Such a case is realized, for example, in the geometry of 
a long and narrow channel at zero magnetic field \cite{soffer}, \cite{dejong}. 
The case of fully specular boundary scattering corresponds to $P_{\varepsilon}(
\varphi, \varphi')=0$ so that ${\rm r}_{\varepsilon \varphi} =1$. 
A fully diffusive boundary scattering means ${\rm r}_{\varepsilon \varphi} =0$ 
(except for the angles $\varphi=0$ and $\varphi=\pi$) and Eq. (2) takes the 
form \cite{beenakker}
\begin{eqnarray}
f_{\varepsilon \varphi}= \frac{1}{2} \int_{0}^{\pi} d \varphi' 
\sin \varphi' f_{\varepsilon 2 \pi-\varphi'}, ~ 0 < \varphi < \pi.
\end{eqnarray}
The function $P_{\varepsilon}(\varphi,\varphi')$ can be representable as a 
product of two functions of $\varphi$ and $\varphi'$, so the kernel in 
Eq. (2) is degenerate. Physically, this case corresponds to uncorrelated boundary 
scattering, when the scattering probability does not depend on the difference 
between the momenta of incoming and reflected particles. The boundary condition 
then can be written in terms of the reflection coefficient 
${\rm r}_{\varepsilon \varphi}$ only:
\begin{eqnarray}
f_{\varepsilon \varphi} = {\rm r}_{\varepsilon \varphi} f_{\varepsilon 
2 \pi-\varphi} + (1-{\rm r}_{\varepsilon \varphi}) M, ~ \varphi \in [0,\pi],
\end{eqnarray}
where $M$ is a constant,
\begin{eqnarray}
M= \frac{1}{{\cal N}} \int_{0}^{\pi} d \varphi \sin \varphi (1-{\rm r}_{
\varepsilon \varphi}) f_{\varepsilon 2 \pi-\varphi}, \\
{\cal N} = \int_{0}^{\pi} d \varphi \sin \varphi (1-{\rm r}_{\varepsilon \varphi}).
\nonumber 
\end{eqnarray}
Naturally, the limiting transition 
${\rm r}_{\varepsilon \varphi} \rightarrow 0$ transforms Eq. (4) into Eq. (3). 
The boundary condition Eq. (4) will be applied below in the calculations.

\begin{figure}[ht!]
\includegraphics[width=8.5cm,clip=]{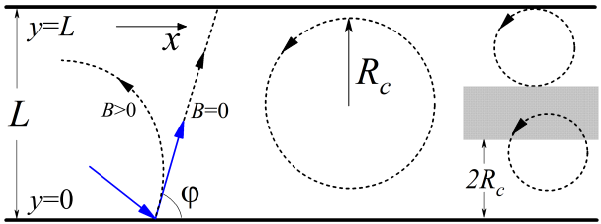}
\caption{\label{fig.1}(Color online) Illustration of electron motion in the 2D channel. The 
diffusive boundary scattering of electrons increases the resistance. A deflection of ballistic 
electron paths (dashed lines) by magnetic field decreases the probability of boundary scattering. 
At $R_c < L/2$, some electrons are moving in cyclotron orbits and do not hit the boundaries. 
At $R_c < L/4$, all the electrons whose ballistic paths pass through the region 
$2R_c < y < L-2R_c$ (shaded) do not hit the boundaries.}
\end{figure}

In this paper, we consider infinitely long 2D channels of width $L$ ($0<y<L$, $-\infty< x 
< \infty$), Fig. 1. Such a model can be applied to samples whose length is much 
larger than their width. In these conditions, the electron system is homogeneous 
along the $x$ direction so that the distribution function depends only on the 
$y$ coordinate, and the electrostatic potential is representable in the form 
$\Phi({\bf r})= -Ex + \Phi(y)$, where $E \equiv E_x$ is a homogeneous electric 
field. Considering the linear response problem, it is convenient to write the 
distribution function as 
\begin{eqnarray}
f_{\varepsilon \varphi}({\bf r})= f_{\varepsilon} - \frac{ \partial f_{\varepsilon}}{\partial 
\varepsilon} [g_{\varepsilon \varphi}(y)- e \Phi(y)],
\end{eqnarray}
where $f_{\varepsilon}$ is the equlibrium Fermi distribution and $g_{\varepsilon \varphi}$ 
describes a small non-equilibrium part of the distribution function. Substituting 
Eq. (6) into Eq. (1), one gets the linearized kinetic equation for $g_{\varepsilon \varphi}$:
\begin{eqnarray}
\left[ \sin{\varphi} \frac{\partial}{\partial y} g_{\varepsilon \varphi}(y) +
R^{-1}_{c \varepsilon} \frac{\partial}{\partial \varphi} g_{\varepsilon \varphi}(y) 
-eE \cos \varphi \right] \nonumber \\
\times \left( \frac{ \partial f_{\varepsilon}}{\partial 
\varepsilon} \right) + \frac{{\cal J}_{\varepsilon \varphi}(y)}{v_{\varepsilon}}=0,
\end{eqnarray}
where $R_{c \varepsilon}$ is the classical cyclotron radius for an electron with 
energy $\varepsilon$ and 
$$
{\cal J}_{\varepsilon \varphi}(y)= - \frac{\partial f_{\varepsilon}}{\partial \varepsilon} 
\left[ J^{im}_{\varepsilon \varphi}(y)+J^{ph}_{\varepsilon \varphi}(y)+
J^{ee}_{\varepsilon \varphi}(y) \right]
$$
is the linearized collision integral describing 
interaction of electrons with impurities (im) and phonons (ph) as well as electron-electron 
(ee) interaction. In the transformations, we have used the equality $E_y({\bf r})= - 
\partial \Phi(y)/\partial y$ and divided the kinetic equation by the velocity 
$v_{\varepsilon}$. It is easy to observe that $g_{\varepsilon \varphi}$ is governed by 
the same boundary condition, Eq. (4), since any angular-independent part of 
$f_{\varepsilon \varphi}({\bf r})$ satisfies Eq. (4) automatically. 

Further, we consider the case of degenerate electron gas, which means that the factor
$-\left( \partial f_{\varepsilon}/\partial \varepsilon \right)$ in Eq. (7) represents 
a narrow peak around the Fermi energy $\varepsilon_F$. Assuming that scattering times 
in the bulk and the boundary reflection coefficients do not change appreciably within 
the temperature-size energy interval around $\varepsilon_F$, one can replace 
$-\left( \partial f_{\varepsilon}/\partial \varepsilon \right)$ by the delta-function 
$\delta(\varepsilon-\varepsilon_F)$ and integrate Eq. (7) over energy, which is 
equivalent to substitution of $\varepsilon$ by $\varepsilon_F$, so the energy index 
below will be omitted. The relative corrections to the resistance caused by the 
thermal broadening of the Fermi distribution are of the order $(T/\varepsilon_F)^2$
and, therefore, are not significant. The electron-electron part of the linearized 
collision integral is written in the relaxation-time approximation \cite{dejong}, 
\cite{govorov}, \cite{scaffidi}, \cite{lucas1}, \cite{lucas2}, \cite{holder}:
\begin{eqnarray} 
J^{ee}_{\varphi}(y) =  
-\frac{g_{\varphi}(y)-g_0(y)-g_1(y) \cos \varphi-\tilde{g}_1(y) \sin \varphi}{\tau_e}, 
\end{eqnarray} 
where $\tau_e$ is the effective electron-electron scattering time, and
\begin{eqnarray} 
g_0= \overline{g_{\varphi}},~g_1= 2 \overline{g_{\varphi} \cos \varphi },~\tilde{g}_1= 
2 \overline{g_{\varphi} \sin \varphi}.
\end{eqnarray} 
Here, $\overline{F}_{\varphi} \equiv (2 \pi)^{-1} \int_0^{2 \pi} d \varphi  F_{\varphi}$ 
denotes angular averaging. The quantities $g_1(y)$ and $\tilde{g}_1(y)$ are proportional 
to local electric currents along $x$ and $y$ directions. Note, however, that in the 
geometry under consideration the current flows only in the $x$ direction, because only 
in this case the requirement of zero flux through the boundary is compatible with the 
continuity equation, so $\tilde{g}_1(y)=0$. A similar relaxation-time approximation 
is applied for the momentum changing (electron-impurity and electron-phonon) parts 
of the collision integral: 
\begin{eqnarray} 
J^{im}_{\varphi}(y) + J^{ph}_{\varphi}(y)= - \frac{g_{\varphi}(y)-g_0(y)}{\tau_{tr}},
\end{eqnarray} 
where $\tau_{tr}$ is the transport time. The times $\tau_{tr}$ and $\tau_e$ characterize 
relaxation of non-equilibrium distribution over the angle of electron momentum. As follows 
from Eqs. (8) and (10), $\tau_{tr}$ describes relaxation of all angular harmonics of the 
distribution function except the zero one ($g_0$), while $\tau_e$ describes relaxation of 
all angular harmonics except the zero and the first ones. Though the introduction of the 
unified times for all harmonics is a crude approximation, it enormously simplifies solution 
of the kinetic equation.
  
Combining Eqs. (7), (8), and (10), we introduce characteristic mean free path lengths 
$l_1=v \tau_{tr}$, $l_e=v \tau_{e}$, and $l=(1/l_1+1/l_e)^{-1}$, and write the 
linearized kinetic equation in the form
\begin{eqnarray}
\left[ \sin{\varphi} \frac{\partial}{\partial y} +
R_c^{-1} \frac{\partial}{\partial \varphi} + \frac{1}{l} \right] g_{\varphi}(y) \nonumber \\
= \frac{g_0(y)}{l}  + \frac{g_1(y) \cos \varphi}{l_e} + e E \cos \varphi 
\equiv {\cal F}_{\varphi}(y).
\end{eqnarray}
This partial differential equation describes the distribution function in the channel
$0<y<L$ with the boundary conditions [see Eq. (4)] written below for $\varphi \in [0,\pi]$:
\begin{eqnarray}
g_{\varphi}(0)= {\rm r}^{0}_{\varphi} g_{2 \pi -\varphi}(0) + (1-{\rm r}^{0}_{\varphi}) M_0, \\   
g_{2\pi-\varphi}(L)={\rm r}^{L}_{\varphi} g_{\varphi}(L) +(1-{\rm r}^{L}_{\varphi}) M_L.
\end{eqnarray}
The two boundaries, in general, can be different, so they are characterized by different 
reflection coefficients, ${\rm r}^{0}_{\varphi}$ for $y=0$ and ${\rm r}^{L}_{\varphi}$ 
for $y=L$. The constants in Eqs. (12) and (13) are
\begin{eqnarray}
M_0= \frac{1}{{\cal N}_0} \int_{0}^{\pi} d \varphi \sin \varphi (1-{\rm r}^{0}_{\varphi}) 
g_{2 \pi-\varphi}(0), \nonumber \\
M_L= \frac{1}{{\cal N}_L} \int_{0}^{\pi} d \varphi \sin \varphi (1-{\rm r}^{L}_{\varphi}) 
g_{\varphi}(L), \nonumber \\
{\cal N}_{0,L}= \int_{0}^{\pi} d \varphi \sin \varphi (1-{\rm r}^{0,L}_{\varphi}). 
\end{eqnarray}
The cyclotron radius at the Fermi level is determined by the magnetic field and electron 
density $n_s$, since $R_c=\ell^2 k_F$, where $\ell=\sqrt{\hbar c/|e|B}$ is the magnetic 
length, $k_F=\sqrt{4 \pi n_s/{\rm g}}$ is the Fermi wavenumber and ${\rm g}$ is the band 
degeneracy factor (${\rm g}=2$ for GaAs quantum wells). Thus, Eqs. (11)-(14) do not contain 
parameters related to band dispersion and can be applied to any kind of fermions, including 
electrons in graphene (where ${\rm g}=4$ due to both spin and valley degeneracy). 

The problem described by Eqs. (11)-(14) is solved by the method of characteristics
as described in the Appendix. Such a solution allows us to reduce the problem to a pair of 
coupled Fredholm integral equations for the functions of one variable, $g_0(y)$ and 
$g_1(y)$: 
\begin{eqnarray}
g_0(y)= eE {\cal L}_{0}(y) + \frac{1}{l} \int_{0}^{L} d y'{\cal K}_{00}(y,y') g_0(y') \nonumber \\
+ \frac{1}{l_e} \int_{0}^{L} d y' {\cal K}_{01}(y,y') g_1(y'), \\
g_1(y)= eE {\cal L}_{1}(y) 
+ \frac{1}{l} \int_{0}^{L} d y' {\cal K}_{10}(y,y') g_0(y')  \nonumber \\ 
+ \frac{1}{l_e} \int_{0}^{L} d y' {\cal K}_{11}(y,y') g_1(y'), 
\end{eqnarray}
where the four kernels ${\cal K}_{nn'}$ and the functions ${\cal L}_{n}$ are given in 
the Appendix. If electron-electron interaction is neglected, $l_e \rightarrow \infty$,  
the terms with ${\cal K}_{01}$ and ${\cal K}_{11}$ disappear, and the first equation 
decouples from the second one. In this limit, the theory describes a classical size 
effect without viscosity corrections. In the limit $B=0$, the terms with ${\cal K}_{01}$, 
${\cal K}_{10}$, and ${\cal L}_{0}$ disappear so that $g_0(y)=0$ and only one integral 
equation remains:
\begin{eqnarray}
g_1(y)= eE {\cal L}_{1}(y) + \frac{1}{l_e} \int_{0}^{L} d y' {\cal K}_{11}(y,y') g_1(y'). 
\end{eqnarray}
This equation is identical to the one derived in Ref. \cite{dejong}, see the Appendix for 
details. It describes effects of viscosity on the transport at zero magnetic field.

A numerical solution of Eqs. (15) and (16) determines $g_0(y)$ and $g_1(y)$ as a 
response to the electric field $E$. Such a solution is obtained by a direct application 
of linear algebra (200-point discretization of the variable $y/L$ is sufficient in most 
cases). A solution by the method of iterations gives the same output. To control the 
accuracy of the procedure, the quantity ${\tilde g}_{1}(y)=2 \overline{g_{\varphi}(y) 
\sin \varphi}$, which is proportional to the current along the $y$ axis and must be zero, 
is calculated simultaneously. At the edges $y=0$ and $y=L$, ${\tilde g}_{1}(y)$ is exactly 
zero, as dictated by the boundary conditions, while in the bulk it is finite because of 
computational errors, but always stays several orders of magnitude smaller than $g_1(y)$. 

The quantity $g_1(y)$, as already noted, describes spatial distribution of electric current 
density $j(y)$. On the other hand, the quantity $g_0(y)$ describes spatial distribution of 
the electrochemical potential, i.e., the local voltage $V(y)$. To show this, we note 
that the latter is defined as $V(y)=\Phi(y)+\delta \mu(y)/e$, where $\delta \mu$ 
is the non-equilibrium part of the local chemical potential. By definition, $\delta \mu(y) = 
\delta n_s (y)/\rho_{2D}$, where $\delta n_s$ is the non-equilibrium part of local electron 
density and $\rho_{2D}=m/\pi \hbar^2$ is the density of states for 2D electrons. 
Thus, $\delta \mu(y)=\int d \varepsilon (\overline{f_{\varepsilon \varphi}(y)}-
f_{\varepsilon}) \simeq g_{0}(y)- e \Phi(y)$, according to Eq. (6). In summary,
\begin{eqnarray}
j(y)= e m v g_1(y)/2 \pi \hbar^2,~~~V(y)=g_0(y)/e.
\end{eqnarray}
These two variables is all we need to find both the longitudinal and the Hall 
resistance. Though the presence of electric field $E$ along the channel induces $y$-dependent 
electrostatic potential $\Phi(y)$ and non-equilibrium part of electron density $\delta n_s (y)$, 
which can be determined by involving the Poisson's equation, we do not need them for description 
of the resistance within the approximations used: the classical transport regime, the linear 
response regime, and the case of degenerate electron gas. 

In the homogeneous case (far away from the boundaries of a wide sample), the solutions of 
Eqs. (15) and (16) are $g_1(y)=eEl_1$ and $g_0(y)=C +eEl_1y/R_c$ (here $C$ is a constant), 
corresponding to the bulk Drude conductivity and constant Hall electric field (see the final 
part of the Appendix for details). 

\begin{figure}[ht!]
\includegraphics[width=8.5cm,clip=]{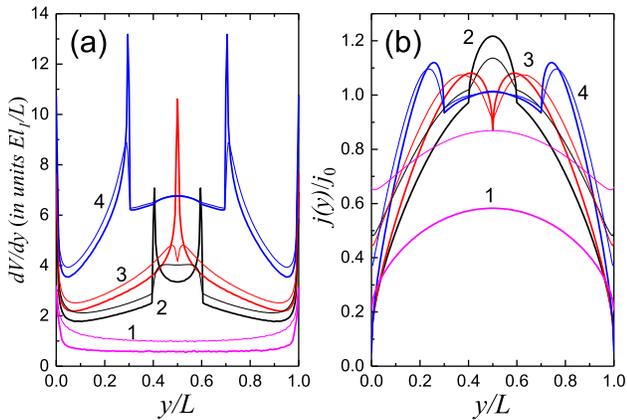}
\caption{\label{fig.2}(Color online) Distribution of the Hall field $dV(y)/dy$ (a) and current density 
(b) at $l_1/L=3$ for several values of magnetic field, $R_c/L=0.6$ (1), 0.3 (2), 0.25 (3), 0.15 (4), in 
the absence of electron-electron scattering. The current density is expressed in units of the 
bulk current density $j_0$.
Sharp features of the distributions associated with ballistic transport appear at $y= 2 R_c$ and 
$y= L - 2 R_c$. The bold lines show the case of fully diffusive boundary scattering, ${\rm r}^{0}_{\varphi} 
={\rm r}^{L}_{\varphi} =0$, while the thin lines correspond to weakly diffusive boundaries, ${\rm r}^{0}_{
\varphi} ={\rm r}^{L}_{\varphi} = \exp(-\alpha \sin^2 \varphi)$ with $\alpha=1$.}
\end{figure}

\begin{figure}[ht!]
\includegraphics[width=8.5cm,clip=]{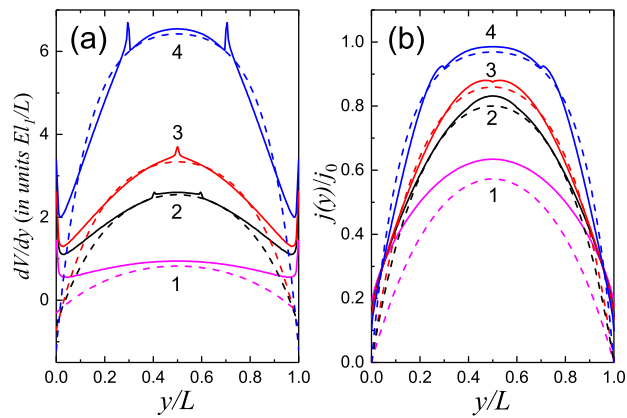}
\caption{\label{fig.3}(Color online) Distribution of the Hall field $dV(y)/dy$ (a) and current density 
(b) at $l_1/L=3$ for several values of magnetic field, $R_c/L=0.6$ (1), 0.3 (2), 0.25 (3), 0.15 (4), when 
the electron-electron scattering is strong, $l_1/l_e=10$. The solid lines 
correspond to calculations for fully diffusive boundary scattering. The dashed lines show the result of application 
of hydrodynamic approximation under the "no-slip" boundary condition, $j(0)=j(L)=0$ ($l_s=0$), see Eq. (19). The 
sharp features of the distributions are suppressed by the viscosity effect.}  
\end{figure}

The examples of calculation of the current and Hall voltage distributions across the 2D channels with a 
high ballisticity, $l_1/L=3$, are shown in Figs. 2 and 3. Instead of $V(y)$, its derivative (Hall field) 
is plotted in order to emphasize sharp features of the distributions appearing at $2R_c < L$ \cite{holder}. 
These features are associated with ballistic motion of electrons in cyclotron orbits. They become weaker with 
increasing specularity of the boundary scattering and tend to disappear when electron-electron interaction 
becomes strong so that the transport enters the hydrodynamic regime \cite{holder}. As shown in Fig. 3, the 
distributions approach the ones calculated in the hydrodynamic approximation \cite{alekseev1}: 
\begin{eqnarray}
j(y)=j_0 \left\{1-\lambda \cosh[ \kappa(y-L/2)] \right\}, \nonumber \\
\frac{dV(y)}{dy} = \frac{E l_1}{R_c} \left\{1-(1+2l/l_1) \lambda \cosh[ \kappa(y-L/2)] \right\}, \\
\kappa=2 \sqrt{\frac{1+(2l/R_c)^2}{l l_1}},\lambda=\frac{1}{\cosh \frac{\kappa L}{2} + 
\kappa l_s \sinh \frac{\kappa L}{2}}, \nonumber
\end{eqnarray}
where $j_0$ is the bulk current density and $l_s$ is the slip length entering the boundary conditions 
$j(y)= \pm l_s \partial j(y)/\partial y$ at $y=0$ and $y=L$. The distributions become closer to the 
hydrodynamic ones as the magnetic field increases. However, near the boundaries the Hall field is still 
considerably different from that following from the hydrodynamic theory. 

When the current and the voltage distributions are found, one can determine the total current 
$I=\int_0^L dy j(y)$ and the Hall voltage $V_H=V(L)-V(0)$ as linear functions of the electric 
field $E$ and to find the longitudinal resistance $R_{xx}$ and the Hall resistance $R_{xy}$. 
A comparison of the results of such calculations to experimental data is described in the next 
section. In Figs. 4-7, we present some results demonstrating the general features 
of the behavior of $R_{xx}$ and $\Delta R_{xy}=R_{xy}-R_{xy}^{(0)}$, expressed in units of 
classical bulk resistances $R_{0}$ and $R_{xy}^{(0)}=B/|e|c n_s$. The magnetic field $B$
is expressed through the ratio $L/R_c \propto B$. We consider the dependence of magnetoresistance 
on the boundary reflection properties, ballisticity ratio $l_1/L$, and relative strength of 
electron-electron scattering $l_1/l_e$. The boundaries are assumed to be equivalent, 
${\rm r}^0_{\varphi}={\rm r}^L_{\varphi} \equiv {\rm r}_{\varphi}$.

\begin{figure}[ht!]
\includegraphics[width=8.5cm,clip=]{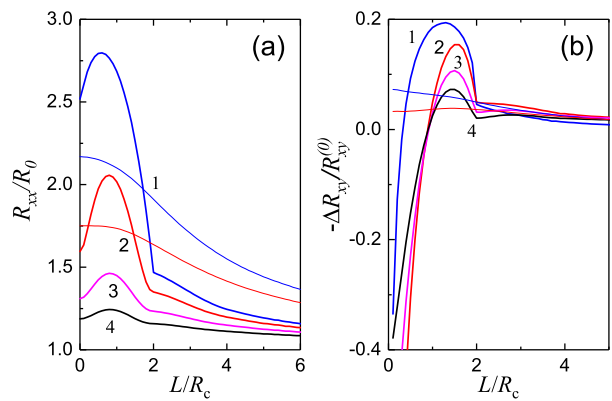}
\caption{\label{fig.4}(Color online) Longitudinal (a) and Hall (b) resistance at $l_1/L=3$ for 
angular-dependent boundary reflection coefficient ${\rm r}_{\varphi} = \exp(-\alpha \sin^2 \varphi)$ 
with $\alpha=\infty$ (fully diffusive, 1), $\alpha=3$ (2), $\alpha=1$ (3), and $\alpha=0.5$ (4). Bold 
lines: $l_1/l_e=0$ (no electron-electron scattering), thin lines (plotted for 1 and 2 only): $l_1/l_e=10$.}   
\end{figure}

\begin{figure}[ht!]
\includegraphics[width=8.5cm,clip=]{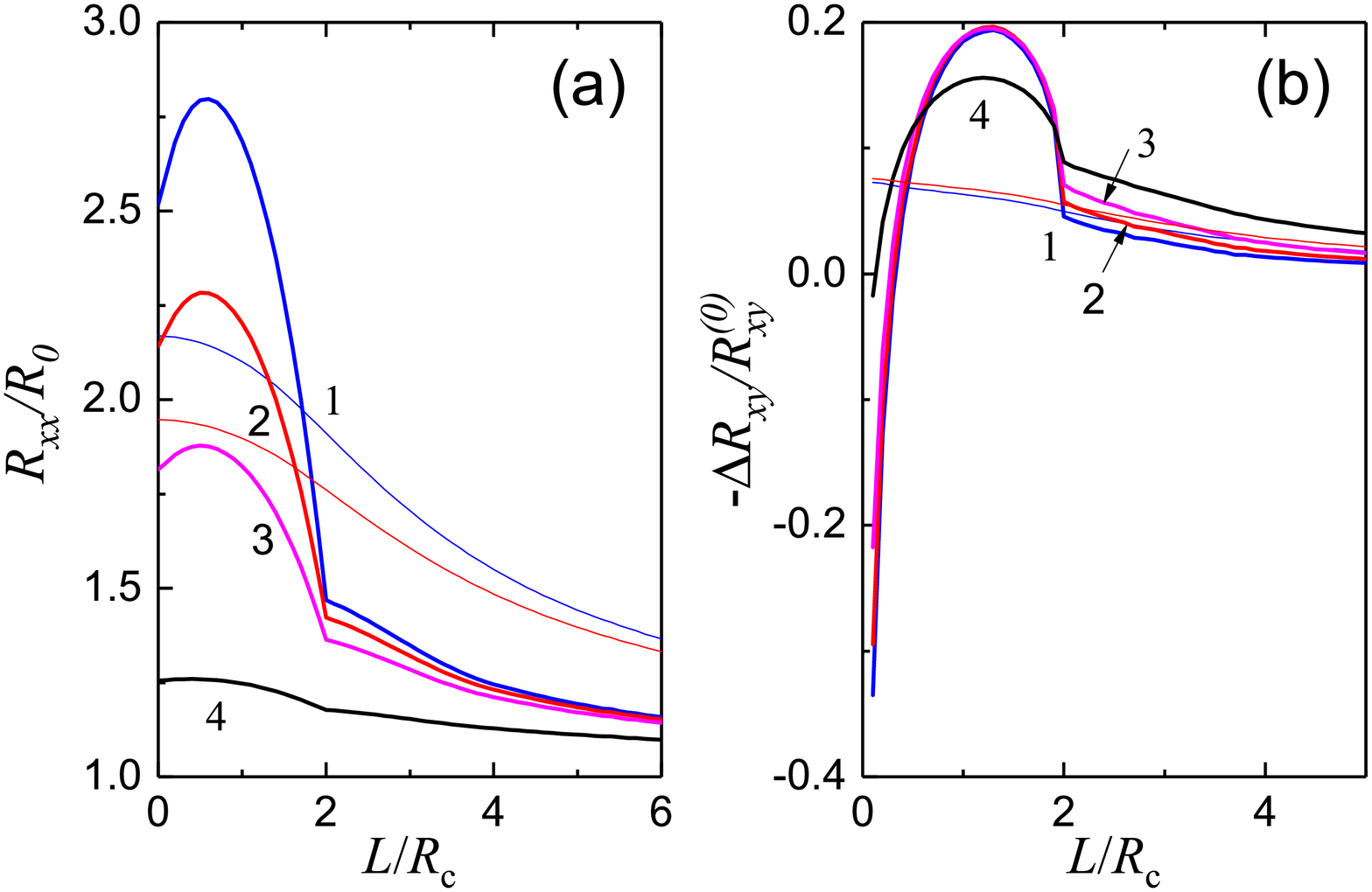}
\caption{\label{fig.5}(Color online) The same as in Fig. 4 for constant (angular-independent) boundary 
reflection coefficient ${\rm r}_{\varphi} = {\rm r}$ with ${\rm r}=0$ (fully diffusive, 1), ${\rm r}=0.2$ 
(2), ${\rm r}=0.4$ (3), and ${\rm r}=0.8$ (4).}   
\end{figure}
   
The basic features of the plots are the manifestations of the classical size 
effect due to quasi-ballistic propagation of 2D electrons in the channel in the presence of a 
magnetic field. They include peaks of both $R_{xx}$ and $-\Delta R_{xy}/R_{xy}^{(0)}$, whose 
maxima are placed at finite magnetic fields, and a sharp decrease of the magnetoresistance slope when 
the cyclotron diameter $2R_c$ becomes smaller than $L$. The behavior of $R_{xx}$ was initially 
described for three-dimensional thin films \cite{ditlefsen} and also observed in submicron-wide 2D 
channels \cite{thornton}, while the behavior of $R_{xy}$ was described recently within the model of 
fully diffusive boundary scattering \cite{scaffidi}. At small $B$, the resistance increases because the 
magnetic field deflects the electrons which move at sliding angles ($\varphi$ close to $0$ or $\pi$) and 
provide a significant contribution to the current. A further increase of $B$, on the contrary, decreases 
the probability of electron collisions with the boundaries, thereby leading to a rapid decrease of the 
resistance. When $2 R_c$ becomes smaller than $L$, there appear electrons which do not collide with 
boundaries during their cyclotron motion, while the electrons scattered by one boundary cannot reach 
the other one unless they are scattered in the bulk. As a result, the decrease of the resistance with 
$B$ slows down considerably. 

\begin{figure}[ht!]
\includegraphics[width=8.5cm,clip=]{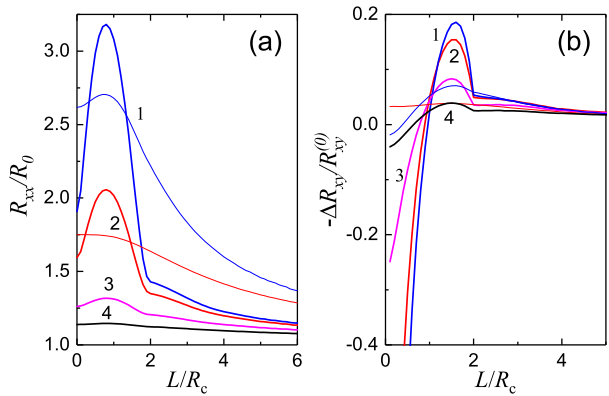}
\caption{\label{fig.6}(Color online) Longitudinal (a) and Hall (b) resistance for the case of 
angular-dependent boundary reflection with $\alpha=3$: $l_1/L=6$ (1), 3 (2), 1 (3) and 0.5 (4). 
Bold lines: $l_1/l_e=0$ (no electron-electron scattering), thin lines (plotted for 1 and 2 only): 
$l_1/l_e=10$.}   
\end{figure}   

\begin{figure}[ht!]
\includegraphics[width=8.5cm,clip=]{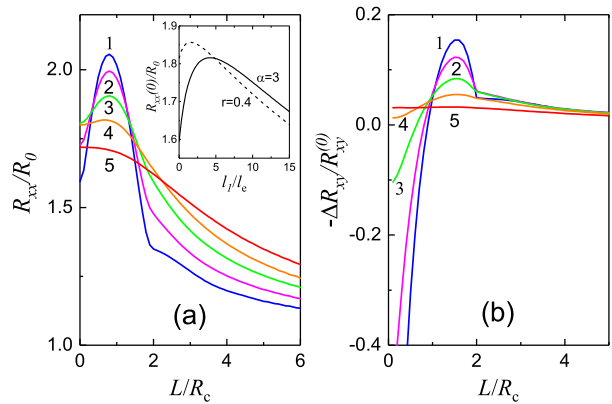}
\caption{\label{fig.7}(Color online) Longitudinal (a) and Hall (b) resistance at $l_1/L=3$, 
for angular-dependent boundary reflection with $\alpha=3$: $l_1/l_e=0$ (1), 1 (2), 3 (3), 6 (4), 
and 12 (5). The inset shows the resistance at $B=0$ vs $l_1/l_e$ for angular-dependent boundary 
reflection with $\alpha=3$ (solid) and for angular-independent boundary reflection with ${\rm r}=0.4$ 
(dash).}   
\end{figure}

Figures 4 and 5 correspond to two different models of boundary reflection. They show 
a decrease of the resistance peaks as the specularity increases. The model of angular-dependent 
boundary reflection, ${\rm r}_{\varphi} = \exp(-\alpha \sin^2 \varphi)$ \cite{soffer}, gives deeper 
local minima of both $R_{xx}$ and $-\Delta R_{xy}/R_{xy}^{(0)}$ 
at $B=0$ because it provides larger probabilities of specular scattering at sliding angles.
In the model of angular-independent reflection, the Hall resistance at $2R_c > L$ is almost 
insensitive to ${\rm r}$ in the region ${\rm r} < 0.5$, if electron-electron scattering is 
absent. Figure 6 demonstrates a rapid decrease of the resistance 
peaks when the ballisticity ratio $l_1/L$ goes down. The increasing specularity and decreasing 
ballisticity suppress the peaks but do not lead to broadening of these peaks and do not remove the 
local minimum at $B=0$. On the other hand, the increase in electron-electron scattering probability, 
which takes place with increasing temperature, not only decreases the height of the peaks, but also 
considerably increases the peak width and leads to a weakening and eventual disappearance of the local 
minimum at $B=0$. Notice also that the peak of $-\Delta R_{xy}/R_{xy}^{(0)}$ is suppressed more rapidly
than the peak of $R_{xx}$. This influence is shown in detail in Fig. 7, which also demonstrates a 
non-monotonic dependence of zero-$B$ longitudinal resistance on $l_1/l_e$. As the electron-electron 
scattering increases, the electron system shifts towards the hydrodynamic regime, when the Gurzhi 
effect \cite{gurzhi} is possible at $B=0$ and the dependence of $R_{xx}$ on $B$ correlates with 
the corresponding dependence of the kinematic viscosity \cite{alekseev1}. Thus, one can say that 
the modifications of the resistance shown in Fig. 7 are manifestations of viscosity effects; see 
also similar results \cite{scaffidi} obtained within the model of fully diffusive boundary scattering. 
Our experimental data are in a good agreement with the behavior discussed above, as presented in 
more detail in the following section.

\section{Comparison of theory with experiment}

We have investigated several samples in the form of long mesoscopic Hall bars of several micron 
widths with 8 symmetrically placed voltage probes (see the inset in Fig. 8). The samples were
fabricated from high-quality GaAs quantum wells with a width of 14 nm. The measurements were 
carried out in a VTI cryostat, using a conventional lock-in technique to measure the resistances 
with a sufficiently low ac current of $0.1-1.0$ $\mu$A passed through contacts 1 and 6. Figure
8 shows a series of plots of longitudinal resistance versus magnetic field $B$ in the region 
of small $B$, where classical magnetotransport is expected. The resistance is measured between 
contacts 4 and 5 (the distance between the centers of the corresponding side arms of the Hall bar 
is 9 $\mu$m, the width of the side arms is 3 $\mu$m at the entry to the channel) in the sample 
with the channel width $L=5$ $\mu$m, electron density $n_s=6.6 \times 10^{11}$ cm$^{-2}$, and 
mobility $2.1 \times 10^{6}$ cm$^2$/V s at $T=4.2$ K. The density remains constant in the range 
of temperatures studied, and the corresponding Fermi energy, wavenumber, and velocity are 
23.6 meV, 0.20 nm$^{-1}$, and $3.52 \times 10^7$ cm/s. The temperature dependence of resistance at $B=0$ 
in macroscopic 2D samples (before shaping the mesoscopic Hall bars) was linear, $R \propto 1 + \beta T$, with 
$\beta \simeq 0.09$ K$^{-1}$ above $4.2$ K, due to the contribution of electron-phonon scattering into the 
transport. For these parameters, a high ballisticity is achieved, when the mean free path $l_1$ is 
larger than $L$ even at $T \simeq 30$ K. Figure 9 shows the temperature dependence 
of $l_1$ for this sample and also for another sample described below. 

\begin{figure}[ht!]
\includegraphics[width=8.5cm,clip=]{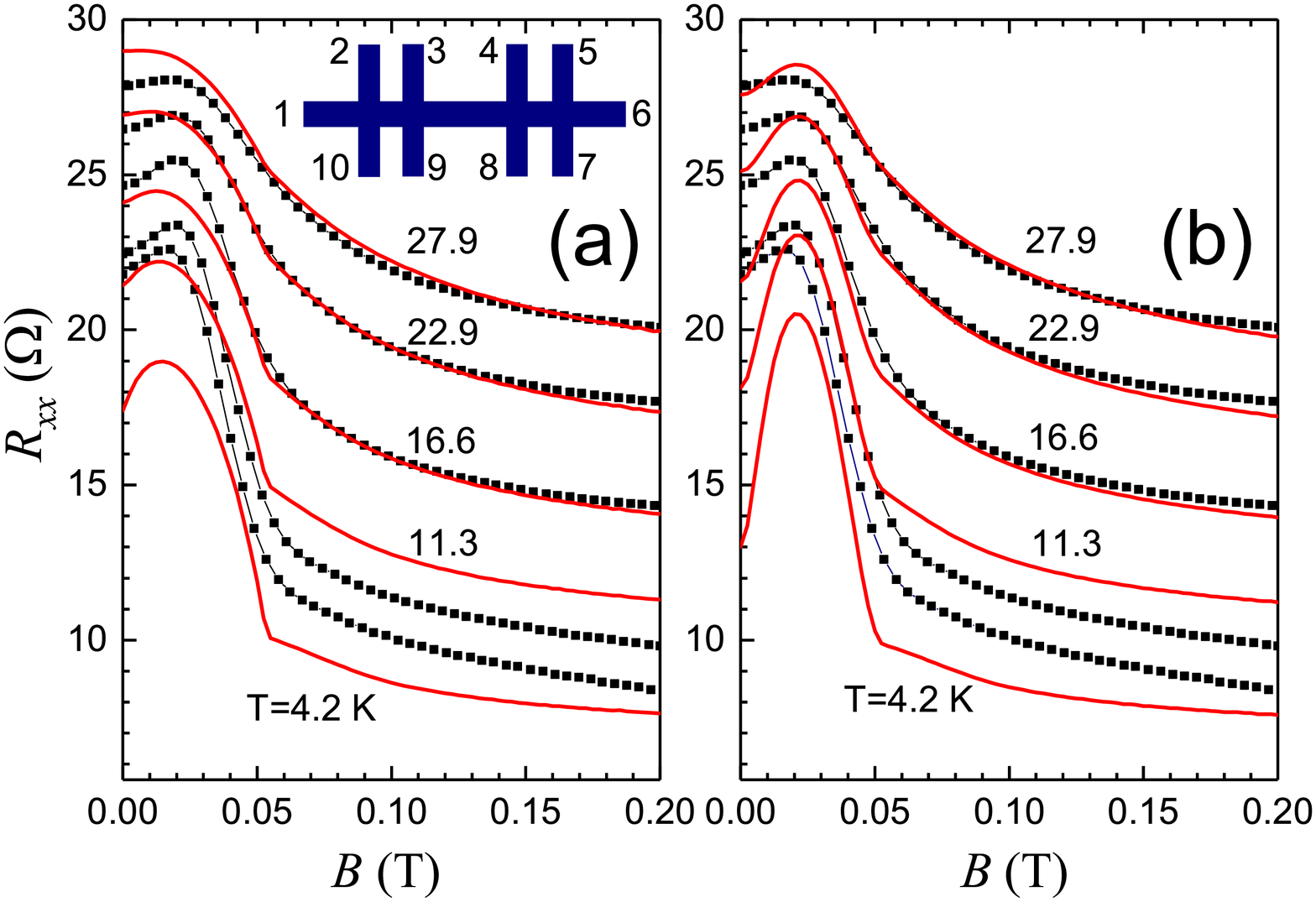}
\caption{\label{fig.8}(Color online) Experimental (points) and calculated (lines) longitudinal 
resistance in the 5 $\mu$m-wide mesoscopic Hall bar (see the parameters in the text), measured 
between contacts 4 and 5 at different temperatures $T=4.2$, 11.3, 16.6, 22.9, and 27.9 K. 
Results of calculations in the parts (a) and (b) correspond to two different models of boundary 
reflection, ${\rm r}_{\varphi}={\rm r}=0.35$ and ${\rm r}_{\varphi} = 
\exp(-\alpha \sin^2 \varphi)$ with $\alpha=3$.}
\end{figure}

All the experimental plots show characteristic peaks in the region of small $B$. An abrupt decrease 
of the peak slope at low temperatures occurs near $B \simeq 0.05$ T, which corresponds to $2R_c=L$ 
($B = 0.053$ T). With increasing $T$, the relative height of the peak becomes smaller and the peak 
width increases. The maximum of the peak is placed at $B \simeq 0.02$ T. The local minimum at $B=0$ 
tends to disappear at high temperatures. The observed weakening of the local minimum at $B=0$, the 
decrease of the relative height of the peak, and the increase of the peak width with increasing $T$ 
cannot be explained solely by a decrease of the transport mean free path length $l_1$ with increasing 
$T$. The contribution of electron-electron scattering turns out to be crucially important for 
description of the experiment. 

\begin{figure}[ht!]
\includegraphics[width=8.5cm,clip=]{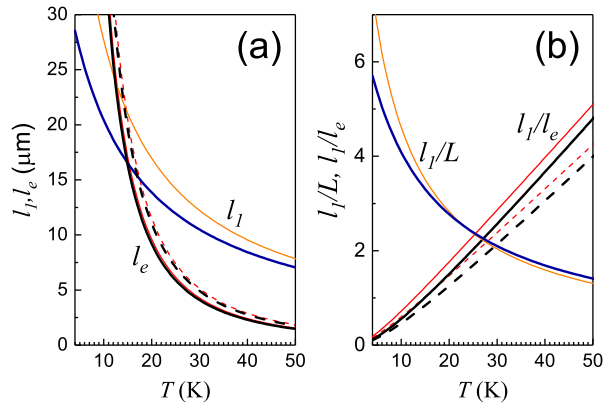}
\caption{\label{fig.9}(Color online) Dependence of characteristic lengths $l_1$ and $l_e$ (a) and of 
the ratios $l_1/L$ and $l_1/l_e$ (b) on temperature for the samples with $L=5$ $\mu$m (bold lines) and 
$L=6$ $\mu$m (thin lines). The length $l_1$ is extracted from the experimental dependence of bulk zero-$B$ 
resistance on temperature, while the length $l_e$ is evaluated according to Eq. (20) (solid and dashed 
lines correspond to $A=5$ and $A=6$, respectively).}   
\end{figure}

Figure 8 also shows the results of calculations based on the parameters ($L$, $n_s$, mobility, 
aspect ratio, and coefficient $\beta$) of the sample described above. The boundaries are 
assumed to be equivalent, ${\rm r}^0_{\varphi}={\rm r}^L_{\varphi} \equiv {\rm r}_{\varphi}$. 
The plots in Fig. 8 (a) and Fig. 8 (b) differ only by the model of boundary reflection. We have 
applied the models of a constant reflection coefficient ${\rm r}_{\varphi} = {\rm r}$ (a) and 
the angular-dependent one, in the form ${\rm r}_{\varphi} = \exp(-\alpha \sin^2 \varphi)$ (b). 
The values of ${\rm r}$ and $\alpha$ have been considered as fitting parameters. The temperature 
dependence of the effective time of electron-electron scattering has been described by the formula 
\begin{eqnarray}
\tau_e= A \frac{\hbar \varepsilon_F}{T^2}. 
\end{eqnarray} 
We emphasize that $\tau_e$, according to its introduction in Eq. (8), is the time 
of relaxation of electron distribution over the angles of electron momentum, and it is different 
from the quantum lifetime of electrons with respect to electron-electron scattering, though follows 
the same $T^{-2}$ dependence. The numerical constant $A$ is treated as another fitting parameter. 
The two fitting parameters, ${\rm r}$ and $A$ (or $\alpha$ and $A$), have been varied to describe 
the heights and the shapes of the magnetoresistance peaks for the entire family of magnetoresistance 
curves plotted at different temperatures. The aspect ratio $L_x/L$, which is a constant scaling 
factor for all $R_{xx}$ plots, was also adjusted by varying the effective distance $L_x$ between 
the side arms of the Hall bar within the interval of the width of these arms, with the best 
fit for $L_x=7.5$ $\mu$m. Since the magnetoresistance peak width is sensitive to $\tau_e$ and 
almost insensitive to ${\rm r}$ or $\alpha$, such fits allow one to estimate the value of $A$ with a good 
accuracy. The best fits are achieved for reasonable values ${\rm r}=0.35$ and $\alpha=3$, with $A=6$ 
for constant reflectivity and $A=5$ for angular-dependent reflectivity. Decreasing $A$ (i.e., 
increasing the contribution of electron-electron scattering) below these values leads to 
broader peaks and, consequently, to a worse agreement with the experiment at high $T$. An 
example of variation of ${\rm r}$ and $A$ is shown in Fig. 10 (for the other sample) by the 
dashed lines.

In the region of low temperatures, $T < 12$ K, the $T$-dependence of $R_{xx}$ in the mesoscopic 
bars turns out to be slower than that for macroscopic samples. We attribute this effect to 
the mesoscopic nature of the contacts. Indeed, although in theory one can formally define 
the electrochemical potential (i.e., the local voltage) in each point of the 2D channel, it is 
not clear whether the voltage measured at the contact connected to the arm of the mesoscopic 
Hall bar corresponds to the voltage at the edge of the channel, especially when temperature is low.
As a consequence, there are vertical shifts between calculated and experimental plots, 
since the calculations are based on the linear $T$-dependence of $R_{xx}$ obtained 
for macroscopic samples. Such shifts can be eliminated by proper scaling factors. As concerns the 
shape of magnetoresistance curves, the agreement between theory and experiment is reasonably good at all 
temperatures. The model of angular-dependent boundary reflection, which is apparently more realistic, 
gives a better agreement. However, this model strongly overestimates the depth of the observed 
local minima at $B=0$, for which a better agreement is given by the model of constant reflection 
coefficient. The deep local minima similar to those in Fig. 8 (b) have been found in earlier 
experiments on 2D channels with lower mobility and submicron widths \cite{thornton}.

\begin{figure}[ht!]
\includegraphics[width=9cm,clip=]{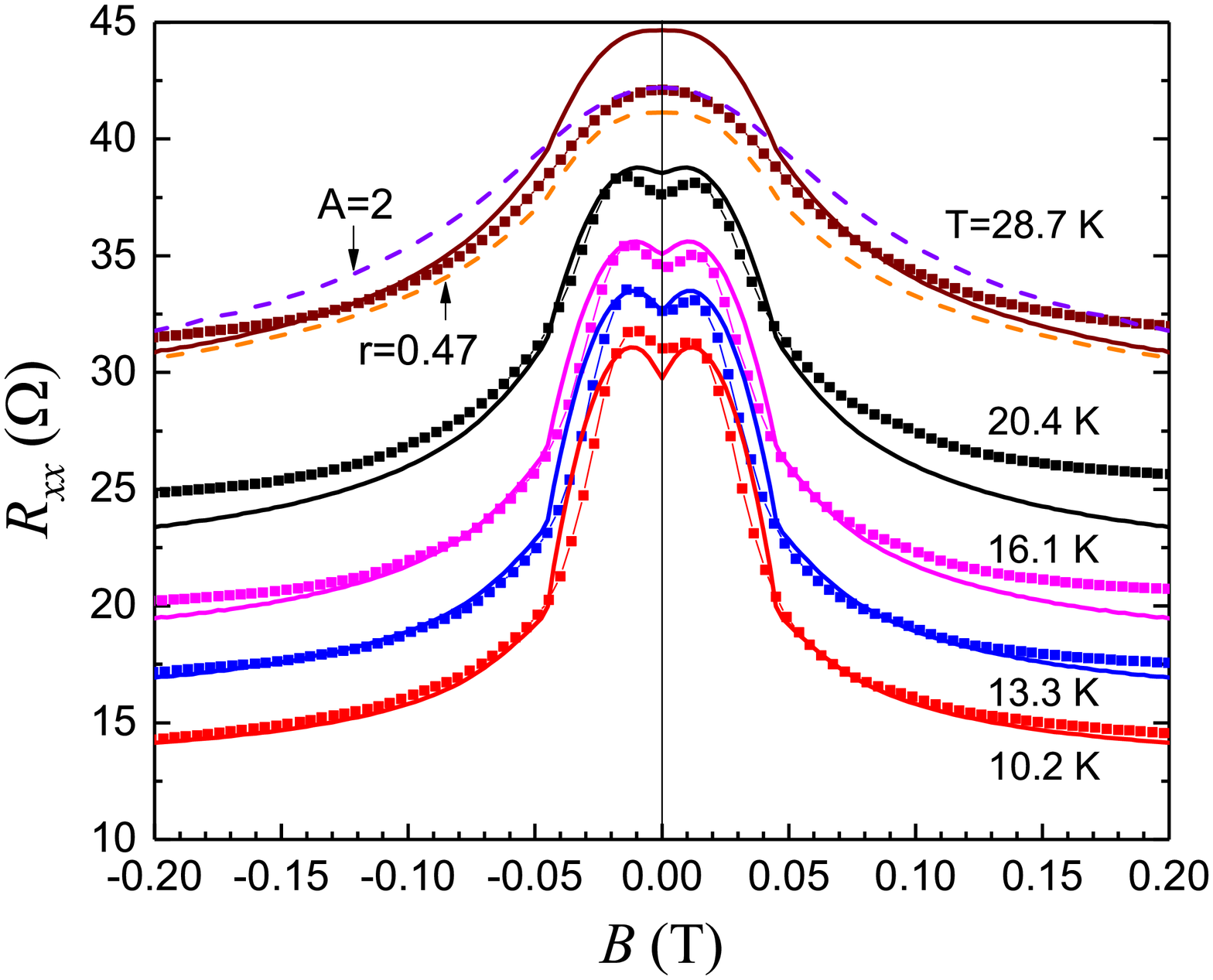}
\caption{\label{fig.10}(Color online) Experimental (points) and calculated (lines) longitudinal 
resistance in the 6 $\mu$m-wide mesoscopic Hall bar (see the geometry in the inset to Fig. 8 and 
the parameters in the text), measured between contacts 3 and 4 at different temperatures 
indicated in the plot. The calculations correspond to the model of constant reflection coefficient, 
with ${\rm r}=0.35$ and $A=6$ [the same as in Fig. 8 (a)]. The dashed lines for $T=28.7$ K are 
calculated with ${\rm r}=0.47$ and $A=6$ and with ${\rm r}=0.35$ and $A=2$.}
\end{figure}

\begin{figure}[ht!]
\includegraphics[width=9cm,clip=]{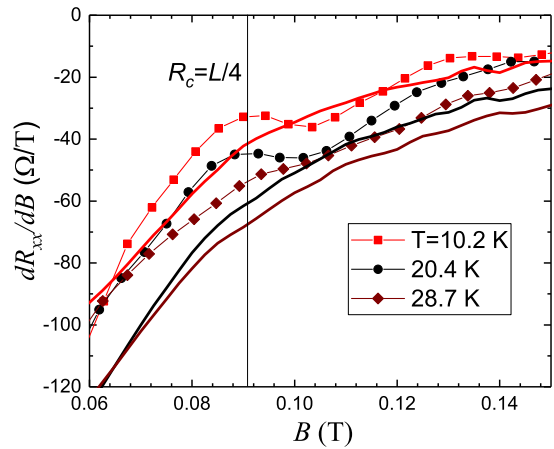}
\caption{\label{fig.11}(Color online) First derivatives of the experimental (points) and calculated 
(lines) magnetoresistance in the 6 $\mu$m-wide mesoscopic Hall bar indicate a modification of transport 
behavior near the point $R_c=L/4$ (vertical line).}
\end{figure}

In addition to the sample described above, we also studied a slightly wider mesoscopic bar, $L=6$ $\mu$m, 
with density $n_s=6.8 \times 10^{11}$ cm$^{-2}$, made from a structure with a higher mobility, $3.2 \times 
10^{6}$ cm$^2$/V s at $T=4.2$ K. For this sample, we have measured the resistance $R_{xx}$ between 
contacts 3 and 4 (see the inset in Fig. 8) separated by $20$ $\mu$m, and also the Hall resistance in the 
region of small $B$. Figure 10 presents the results of measurements together with theoretical plots 
for the model of angular-independent boundary scattering, calculated for the same parameters as those 
used in Fig. 8 (a), ${\rm r}=0.35$ and $A=6$. Again, we have a reasonable agreement between theory and 
experiment. For the high-temperature plot, the agreement can be improved by increasing ${\rm r}$, which 
changes the relative height of the magnetoresistance peak without changing its shape, while a decrease
of $A$ makes the peak broader than the experimental one, see the dashed lines in Fig. 10. 

Apart from the discussed manifestations of size effect, both the experimental and theoretical magnetoresistances 
at low temperatures demonstrate a weak modification of their slopes at $B=0.091$ T, which corresponds to the 
condition $R_c=L/4$. As shown in Figs. 2 and 3, at this particular point both the current and the Hall 
field distributions exhibit sharp cusps at the center of the conducting channel. The cusps of the 
distributions recently became a subject of discussion \cite{holder}, but their connection to 
magnetoresistance has not been examined either theoretically or experimentally. Meanwhile, a 
modification of magnetoresistance is expectable, because at $R_c < L/4$ there opens a region 
$2R_c < y < L-2R_c$ containing the electrons whose ballistic trajectories do not reach any of 
the boundaries, see Fig. 1. To study the behavior of magnetoresistance in the vicinity of $R_c=L/4$, 
we have plotted the first derivatives of $R_{xx}$, shown in Fig. 11. The theoretical plot at 
low $T$ shows a sharp change of the slope at $R_c=L/4$, which is equivalent to a sharp change of 
the second derivative of $R_{xx}$. The experimental plot demonstrates an even stronger feature: 
a change in the slope of $dR_{xx}/dB$ includes an interval of non-monotonic dependence near 
$R_c=L/4$. These features are apparently of the ballistic origin, and they are washed out 
by temperature when the transport approaches to the hydrodynamic regime, as shown by both theory 
and experiment. Similar modifications of the resistance are also present in the case when the 
resistance is measured between contacts 2 and 5 separated by $40$ $\mu$m, but they are not 
seen in the measurements shown in Fig. 8, where the distance between the voltage contacts is 
close to $L$.  

\begin{figure}[ht!]
\includegraphics[width=9cm,clip=]{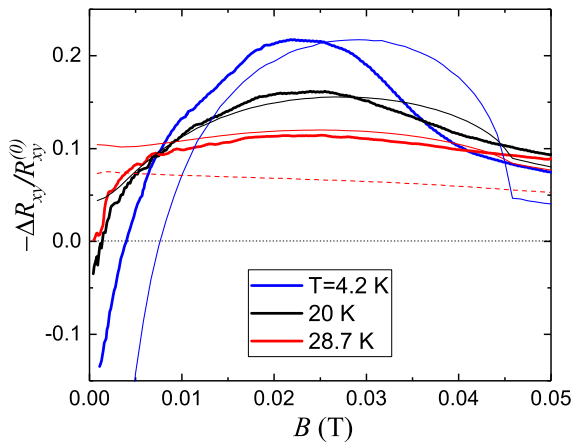}
\caption{\label{fig.12}(Color online) Experimental (bold lines) and calculated (thin lines)
normalized Hall resistance in the 6 $\mu$m-wide mesoscopic Hall bar (see the geometry in the 
inset to Fig. 8 and the parameters in the text) at different temperatures. The calculations 
correspond to the model of constant reflection coefficient, with ${\rm r}=0.35$ and $A=6$, 
the dashed line shows the result for $T=28.7$ K with $A=2$.}
\end{figure}
   
Finally, we describe the results of the Hall resistance measurements shown in Fig. 12.
The theory predicts (see \cite{scaffidi} and the results shown in Figs. 4-7)
that $\Delta R_{xy}$ in the ballistic transport regime changes its sign in the region of low $B$. 
This property has been recently confirmed experimentally \cite{gusev2}, and it is also seen in Fig. 12. 
The comparison of the present theory with experiment shows that the general behavior of the Hall 
resistance and the heights of the peaks of $-\Delta R_{xy}/R_{xy}^{(0)}$ are in agreement with theory. 
However, the experimental peaks are positioned at smaller magnetic field than the theoretical ones. 
We could not obtain a good fit to the shape of $\Delta R_{xy}$ by varying the 
adjustable parameters within the reasonable range. In any case, we find it more reliable to concentrate 
on fitting of $R_{xx}$, since the data of $\Delta R_{xy}$ have a considerably greater measurement 
error compared to $R_{xx}$, because of relative smallness of $\Delta R_{xy}$. 
The non-monotonic experimental and theoretical plots in Fig. 12 indicate that even at $T=28.7$ K the 
hydrodynamic transport regime is not yet reached. Indeed, the hydrodynamic theory describes a monotonic 
decrease of $-\Delta R_{xy}/R_{xy}^{(0)}$ with increasing magnetic field \cite{scaffidi,holder}; in 
particular, based of Eq. (19) one can find 
\begin{eqnarray}
\frac{\Delta R_{xy}}{R_{xy}^{(0)}}=-\frac{(2 l/l_1)}{(\kappa L/2)/\tanh(\kappa L/2) + \kappa^2 L l_s/2 - 1 }.
\end{eqnarray}
This dependence does not fit our experimental data at any slip length $l_s$, which is not surprising, 
since the hydrodynamic regime requires $l_1/l_e \gg 1$ and $l_e/L \ll 1$ while our calculations give 
$l_1/l_e \simeq 2$ and $l_e/L \simeq 1$ at $T=28.7$ K (see Fig. 9). Nevertheless, the temperature-induced effects 
such as a rapid suppression of the peak of $-\Delta R_{xy}/R_{xy}^{(0)}$ and the change of sign of $\Delta R_{xy}$ 
at $B \rightarrow 0$ confirm that the influence of electron viscosity on transport properties is already 
significant. 

\section{Discussion and conclusions}

Whereas the $T^{-2}$ scaling of the effective electron-electron scattering time $\tau_e$ given by Eq. (20) follows 
from the general properties of Fermi liquids, the numerical coefficient $A$ in this dependence is a subject of 
discussion. Our observation of the magnetoresistance behavior and its modification by temperature, together 
with a detailed comparison of experimental data with theory, suggest $A \simeq 5-6$, which, at
first glance, seems to be an unexpectedly large value. Below we demonstrate why $A$ actually can be large. 
For degenerate 2D electron gas, when $(T/\varepsilon_F)^2 \ll 1$, the dominant electron-electron scattering 
events are either "collinear" collisions, when the directions of motion of colliding particles are nearly 
equal and the scattering angle is small, or "head-to-head" collisions, when the directions of colliding 
particles are nearly opposite and the scattering angle is arbitrary. While the collinear collisions 
determine the quantum lifetime of electrons \cite{giuliani}, \cite{zheng}, \cite{jungwirth}, they are not efficient 
in relaxation of the angular distribution of electrons, and the main contribution to the electron-electron 
collision integral comes from the head-to-head collisions \cite{gurzhi1,gurzhi2}. The latter, however, 
can be significantly suppressed for the following reasons. First, the wavenumber transferred in the head-to-head 
collisions is of the order of Fermi wavenumber $k_F$, so when $k_F$ exceeds either the inverse screening length 
$q_0=2/a_B$ (here $a_B$ is the Bohr radius) or the inverse quantum well width $1/a$ \cite{goodnick}, the scattering 
amplitude decreases. This is the case of our high-density samples, where $k_F \simeq 3/a$. 
Second, the effect of Cooper-channel renormalization of the scattering amplitude \cite{aleiner}, applicable 
to head-to-head collisions, can enhance the effective electron-electron scattering time by a logarithmically 
large factor $\ln^2(\varepsilon_F/T)$ \cite{novikov}.

The suppression of electron-electron scattering described above makes it difficult to attain 
the fully hydrodynamic regime in GaAs samples, since an increase of temperature over 40-50 K turns on 
a strong scattering of elecrons by optical phonons. Nevertheless, one may identify the intermediate 
regime, when $l_1 > l_e \sim L$ and the influence of electron-electron interaction on angular 
relaxation of electron distribution, promoting the effects of electronic viscosity, becomes 
considerable. In this regime, which is realized in our samples at $T=20-30$ K, all the features characteristic 
for the quasi-ballistic (low-temperature) transport regime are suppressed with increasing temperature. 
The manifestations of the ballistic transport already described in the previous studies are: the peaks of 
both the longitudinal resistance $R_{xx}$ and normalized Hall resistance $-\Delta R_{xy}/R_{xy}^{(0)}$, the 
sharp change of the slope of these peaks at $R_c=L/2$, the local minimum of $R_{xx}$ at $B=0$, and the negative 
sign of $-\Delta R_{xy}/R_{xy}^{(0)}$ at small $B$. To this list, we have added a previously unnoticed feature, 
the sharp change of the slope of the derivative $dR_{xx}/dB$ at $R_c=L/4$. By combining theory and experiment, 
we have demonstrated that the kinetic equation approach, based on the relaxation-time approximation for the 
electron-electron collision integral, gives a reasonably accurate quantitative description of magnetoresistance 
as the latter evolves with temperature. A comparison of theory and experiment allows one to probe the 
contribution of electron-electron interaction into transport coefficients.

Whereas in the region of small magnetic fields the purely classical approach used above is 
valid, the increase of the magnetic field would lead to the quantum Hall regime \cite{beenakker}. A connection 
between classical magnetotransport in a channel with diffusive boundary reflection and quantum magnetotransport 
in a channel with the quantum Hall edge is not yet established theoretically, although some steps in this 
direction are already taken \cite{stegmann}. This challenging and important problem deserves a proper attention 
in the future studies.\\

The authors acknowledge financial support of this work by FAPESP and
CNPq (Brazilian agencies).  

\begin{appendix}

\section{Equations for potential and current distribution}

This Appendix provides the details of derivation of Eqs. (15) and (16) and 
specifies the functions ${\cal K}_{nn'}(y,y')$ and ${\cal L}_{n}(y)$ standing 
in these equations. Also, the limiting transitions to the cases of zero magnetic 
field, semi-infinite plane, and wide channel are described.  

The general solution of Eq. (11) is written as a sum of the general solution of homogeneous 
equation and a solution of inhomogeneous equation (for brevity, $R_c \equiv R$ below):
\begin{eqnarray}
g_{\varphi}= {\cal D} (u)e^{-p\varphi} + \int_{0}^{\varphi} 
d \varphi' e^{p(\varphi'-\varphi)} R {\cal F}_{\varphi'}(y'), \\
p \equiv R/l,~~u=y+R\cos \varphi,~~ y'=u - R \cos \varphi', \nonumber
\end{eqnarray}
where ${\cal D}(u)$ is an arbitrary function of its argument. To find this function, it is 
necessary to apply the boundary conditions. Before doing this, it is convenient to write 
the solution in the regions $0 < \varphi < \pi$ and $\pi < \varphi < 2 \pi$ separately:
\begin{eqnarray}
g_{\varphi}(y)= {\cal D}_0 (y+R\cos \varphi)e^{-p\varphi} + \int_{\varphi_0}^{\varphi} 
d \varphi' \nonumber \\
\times e^{p(\varphi'-\varphi)} R {\cal F}_{\varphi'}(y'),~~~ 0 < \varphi < \pi, \nonumber \\
g_{\varphi}(y)= {\cal D}_1 (y+R\cos \varphi)e^{-p\varphi} - \int_{\varphi}^{2 \pi-\varphi_0} 
d \varphi' \nonumber \\
\times e^{p(\varphi'-\varphi)} R {\cal F}_{\varphi'}(y'),~~~ \pi <\varphi < 2 \pi.
\end{eqnarray}
The requirement $y' \in [0,L]$ imposes restrictions on the range of $\varphi'$. 
Here we introduce important variables:
\begin{eqnarray}
\varphi_0 = \arccos({\rm min}\{1,\cos \varphi + y/R \}), \nonumber \\
\varphi_L = \arccos({\rm max}\{-1,\cos \varphi + (y-L)/R \}), 
\end{eqnarray}
both of them are functions of $y+R\cos \varphi$. 
If $0 < \varphi < \pi$, then $\varphi_0=\varphi$ at the lower 
boundary, $y=0$, and $\varphi_L=\varphi$ at the upper boundary, 
$y=L$. Inside the sample, $\varphi_0 < \varphi < \varphi_L$. 

Application of the boundary conditions (12) and (13) defines ${\cal D}_0$ and 
${\cal D}_1$. In the region $0 < \varphi < \pi$, the solution takes the 
form
\begin{eqnarray}
g_{\varphi}(y)= (1-{\rm r}^0_{\varphi_0}) M_0 e^{p(\varphi_0-\varphi)}/{\rm d}  \nonumber \\
+ (1-{\rm r}^L_{\varphi_L}) M_L e^{p(2\varphi_0-\varphi-\varphi_L)} {\rm r}^0_{\varphi_0}/{\rm d}  
\nonumber \\
+\int_{\varphi_0}^{\varphi_L} d \varphi' R {\cal F}_{\varphi'}(y') \left\{
[ \theta(\varphi-\varphi')+(1-{\rm d})/{\rm d} ]  \right. \nonumber \\ 
\left. \times e^{p(\varphi'-\varphi)} + {\rm r}^0_{\varphi_0} 
e^{p(2 \varphi_0-\varphi-\varphi')}/{\rm d} \right\}, 
\end{eqnarray}
\begin{eqnarray}
g_{2\pi-\varphi}(y)= 
(1-{\rm r}^L_{\varphi_L}) M_L e^{p(\varphi-\varphi_L)}/{\rm d}  \nonumber \\
+ (1-{\rm r}^0_{\varphi_0}) M_0 e^{p(\varphi+\varphi_0-2\varphi_L)} {\rm r}^L_{\varphi_L} /{\rm d}  
\nonumber \\
+\int_{\varphi_0}^{\varphi_L} d \varphi' R {\cal F}_{\varphi'}(y') \left\{
[ \theta(\varphi'-\varphi)+(1-{\rm d})/{\rm d} ]  \right. \nonumber \\ 
\left. \times e^{p(\varphi-\varphi')} + {\rm r}^L_{\varphi_L} 
e^{p(\varphi+\varphi'-2 \varphi_L)}/{\rm d} \right\}. 
\end{eqnarray}
Transforming the integrals over $\varphi'$ into the integrals over $y'$, one also obtains
\begin{eqnarray}
g_{\varphi}(y)= (1-{\rm r}^0_{\varphi_0}) M_0 e^{p(\varphi_0-\varphi)}/{\rm d}  \nonumber \\
+ (1-{\rm r}^L_{\varphi_L}) M_L e^{p(2\varphi_0-\varphi-\varphi_L)} {\rm r}^0_{\varphi_0}/{\rm d}  
\nonumber \\
+\int_{0}^{L} d y' Q^0_{\varphi}(y,y') {\cal F}_{\varphi'}(y'),
\end{eqnarray}
\begin{eqnarray}
g_{2\pi-\varphi}(y)= 
(1-{\rm r}^L_{\varphi_L}) M_L e^{p(\varphi-\varphi_L)}/{\rm d}  \nonumber \\
+ (1-{\rm r}^0_{\varphi_0}) M_0
e^{p(\varphi+\varphi_0-2\varphi_L)} {\rm r}^L_{\varphi_L}/{\rm d}  \nonumber \\
+\int_{0}^{L} d y' Q^1_{\varphi}(y,y') {\cal F}_{\varphi'}(y').
\end{eqnarray}
In these expressions, 
\begin{eqnarray}
{\rm d} = 1 - {\rm r}^0_{\varphi_0}{\rm r}^L_{\varphi_L}e^{2 p(\varphi_0-\varphi_L)},
\end{eqnarray}
and
\begin{eqnarray}
Q^0_{\varphi}(y,y')= \left\{
[ \theta(\varphi-\varphi')+(1-{\rm d})/{\rm d} ] e^{p(\varphi'-\varphi)}  \right. \nonumber \\ 
\left.  + {\rm r}^0_{\varphi_0} 
e^{p(2 \varphi_0-\varphi-\varphi')}/{\rm d} \right\} \frac{1}{\sin \varphi'} , \\
Q^1_{\varphi}(y,y')= \left\{
[ \theta(\varphi'-\varphi)+(1-{\rm d})/{\rm d} ] e^{p(\varphi-\varphi')}  \right. \nonumber \\ 
\left. + {\rm r}^L_{\varphi_L} 
e^{p(\varphi+\varphi'-2 \varphi_L)}/{\rm d} \right\} \frac{1}{\sin \varphi'}, 
\end{eqnarray}
with
\begin{eqnarray}
\varphi'= \arccos\left[ \cos \varphi +(y-y')/R \right].
\end{eqnarray}
It is implied that the kernels $Q^0$ and $Q^1$ are equal to zero outside the region 
$u-R<y'<u+R$, since only in this region the definition of $\varphi'$ makes sense. Since 
$y-R< u < y+R$, this also means that $Q^0$ and $Q^1$ are nonzero within the interval 
$|y'-y| < 2R$, so that the actual upper and lower limits of the integration over $y'$ are 
$y'_{max}={\rm min} \{L,y+2R \}$ and $y'_{min}={\rm max} \{0,y-2R \}$, respectively.  
The correlation length of $2R$ is characteristic for the case $R < l$. However, if $R > l$, 
the correlation length is of the order of the mean free path length $l$, because $Q^0$ and $Q^1$ 
exponentially decrease with $|y'-y|/l$. 
 
At the boundaries,
\begin{eqnarray}
g_{2\pi-\varphi}(0)= 
\frac{1}{{\rm d}_0} \left[(1-{\rm r}^L_{\varphi_{L0}}) M_L e^{p(\varphi-\varphi_{L0})} \right. \nonumber \\
\left. + (1-{\rm r}^0_{\varphi}) M_0 (1 - {\rm d}_0)/{\rm r}^0_{\varphi} \right]
\nonumber \\
+\int_{0}^{L} d y'  \frac{{\cal F}_{\varphi'_0}(y')}{{\rm d}_0 \sin \varphi'_0} \left[
e^{p(\varphi-\varphi'_0)} +  e^{p(\varphi'_0-\varphi)} \frac{1-{\rm d}_0}{{\rm r}^0_{\varphi}} \right],~~ 
\end{eqnarray}
\begin{eqnarray}
g_{\varphi}(L)= 
\frac{1}{{\rm d}_L} \left[ (1-{\rm r}^0_{\varphi_{0L}}) M_0 e^{p(\varphi_{0L}-\varphi)} \right.  \nonumber \\
\left. + (1-{\rm r}^L_{\varphi}) M_L (1 - {\rm d}_L)/{\rm r}^L_{\varphi} \right]  \nonumber \\
+\int_{0}^{L} d y' \frac{{\cal F}_{\varphi'_L}(y')}{{\rm d}_L \sin \varphi'_L} \left[
e^{p(\varphi'_L-\varphi)} + e^{p(\varphi-\varphi'_L)} \frac{1-{\rm d}_L}{{\rm r}^L_{\varphi}} \right],~~ 
\end{eqnarray}
where $\varphi_{L0}$, $\varphi'_0$, and ${\rm d}_0$ denote $\varphi_{L}$, $\varphi'$, and ${\rm d}$ 
at $y=0$, respectively, while $\varphi_{0L}$, $\varphi'_L$, and ${\rm d}_L$ denote $\varphi_{0}$, 
$\varphi'$, and ${\rm d}$ at $y=L$. The expressions (A12) and (A13) can be used to find
the constants $M_0$ and $M_L$ according to Eq. (14). This leads to the following 
linear equations:
\begin{eqnarray}
({\cal N}_0-\alpha_0) M_0  - \beta_0 M_L = \kappa_0,  \nonumber \\
-\beta_L M_0 + ({\cal N}_L-\alpha_L) M_L = \kappa_L,
\end{eqnarray}
where 
\begin{eqnarray}
\alpha_0= \int_0^{\pi} \frac{d \varphi}{{\rm d}_0} (1-{\rm r}^0_{\varphi})^2 {\rm r}^L_{\varphi_{L0}}
\sin \varphi e^{2p(\varphi-\varphi_{L0})}, \nonumber \\
\alpha_L= \int_0^{\pi} \frac{d \varphi}{{\rm d}_L} (1-{\rm r}^L_{\varphi})^2 {\rm r}^0_{\varphi_{0L}}
\sin \varphi e^{2p(\varphi_{0L}-\varphi)}, \nonumber \\
\beta_0= \int_0^{\pi} \frac{d \varphi}{{\rm d}_0} (1-{\rm r}^0_{\varphi})(1-{\rm r}^L_{\varphi_{L0}})
\sin \varphi e^{p(\varphi-\varphi_{L0})}, \nonumber \\
\beta_L= \int_0^{\pi} \frac{d \varphi}{{\rm d}_L} (1-{\rm r}^L_{\varphi})(1-{\rm r}^0_{\varphi_{0L}}) 
\sin \varphi e^{p(\varphi_{0L}-\varphi)},
\end{eqnarray}
and
\begin{eqnarray}
\kappa_0=\int_0^{\pi} d \varphi (1-{\rm r}^0_{\varphi}) \sin \varphi \int_{0}^{L} d y'  
\frac{{\cal F}_{\varphi'_0}(y')}{{\rm d}_0 \sin \varphi'_0} \nonumber \\
\times \left[
e^{p(\varphi-\varphi'_0)} +  e^{p(\varphi'_0-\varphi)}  (1-{\rm d}_0)/{\rm r}^0_{\varphi} \right], \nonumber \\
\kappa_L=\int_0^{\pi} d \varphi (1-{\rm r}^L_{\varphi}) \sin \varphi \int_{0}^{L} d y'  
\frac{{\cal F}_{\varphi'_L}(y')}{{\rm d}_L \sin \varphi'_L} \nonumber \\
\times \left[
e^{p(\varphi'_L-\varphi)} +  e^{p(\varphi-\varphi'_L)}  (1-{\rm d}_L)/{\rm r}^L_{\varphi} \right].
\end{eqnarray}
If two boundaries are equivalent, the following relations are valid:
\begin{eqnarray} 
\alpha_L=\alpha_0,~~ \beta_L=\beta_0,~~ \kappa_L=-\kappa_0.
\end{eqnarray}
To prove the first two equalities, it is sufficient to substitute $\varphi \rightarrow \pi-\varphi$ 
under the integrals in Eq. (A15). This transformation does not affect ${\rm r}_{\varphi}$ 
and $\sin \varphi$, while leading to $\cos \varphi \rightarrow -\cos \varphi$ and $\varphi_{L0} 
\rightarrow \pi - \varphi_{0L}$ so that $\cos \varphi_{L0} \rightarrow -\cos \varphi_{0L}$. To 
prove that $\kappa_L=-\kappa_0$, one should also substitute $y' \rightarrow L-y'$ under the 
integrals over $y'$ in Eq. (A16), which leads to $\varphi'_0 \rightarrow \pi-\varphi'_L$, and 
to notice that ${\cal F}_{\varphi}(y)=-{\cal F}_{\pi-\varphi}(L-y)$. With ${\cal N}_0={\cal N}_L 
\equiv {\cal N}$, $\alpha_0=\alpha_L \equiv \alpha$, $\beta_0=\beta_L \equiv \beta$, 
$\eta_0 \equiv \eta$, and $\kappa_0 \equiv \kappa$, one obtains
\begin{eqnarray} 
M_0=-M_L= \frac{\eta+\kappa}{{\cal N} - \alpha + \beta}.
\end{eqnarray}
In the general case, one may introduce numerical coefficients 
$Z=({\cal N}_0 - \alpha_0) ({\cal N}_L - \alpha_L)- \beta_0 \beta_L$,
$a_{00}=({\cal N}_L - \alpha_L)/Z$, $a_{L0}=\beta_L/Z$, $a_{0L}=\beta_0/Z$, and 
$a_{LL}=({\cal N}_0 - \alpha_0)/Z$, and then   
\begin{eqnarray} 
M_0=a_{00} \kappa_0 + a_{0L} \kappa_L,~ M_L=a_{L0} \kappa_0 + a_{LL} \kappa_L.
\end{eqnarray}

The solutions presented above lead to the integral equations (15) and (16) with ($n=0,1$, $n'=0,1$)
\begin{eqnarray} 
{\cal K}_{nn'}(y,y') = \int_0^{\pi} \frac{d \varphi}{2 \pi} (2 \cos \varphi)^n (\cos \varphi')^{n'} 
Q^+_{\varphi}(y,y')  \nonumber \\
+[\mu^n_0(y) a_{00} + \mu^n_L(y) a_{L0}] \zeta^{n'}_0(y') ~~~\nonumber \\
+[\mu^n_0(y) a_{0L} + \mu^n_L(y) a_{LL}] \zeta^{n'}_L(y'),~~~  
\end{eqnarray}
\begin{eqnarray} 
{\cal L}_{n}(y) = \int_0^{\pi} \frac{d \varphi}{2 \pi} \int_0^L dy' (2 \cos \varphi)^n \cos \varphi' Q^+_{\varphi}(y,y')
\nonumber \\
+[\mu^n_0(y) a_{00} + \mu^n_L(y) a_{L0}] \int_0^L dy' \zeta^1_0(y') ~~~  \nonumber \\
+[\mu^n_0(y) a_{0L} + \mu^n_L(y) a_{LL}] \int_0^L dy' \zeta^1_L(y'),~~~  
\end{eqnarray}
where $Q^+_{\varphi}(y,y')=Q^0_{\varphi}(y,y')+Q^1_{\varphi}(y,y')$,
\begin{eqnarray} 
\mu^n_0(y)= \int_0^{\pi} d \varphi (2 \cos \varphi)^n
\frac{1-{\rm r}^0_{\varphi_0}}{2 \pi {\rm d}} \nonumber \\
\times \left[ e^{p(\varphi_0-\varphi)} +
{\rm r}^L_{\varphi_L} e^{p(\varphi+\varphi_0-2\varphi_L )} \right], 
\end{eqnarray}
\begin{eqnarray} 
\mu^n_L(y)= \int_0^{\pi} d \varphi (2 \cos \varphi)^n
\frac{1-{\rm r}^L_{\varphi_L}}{2 \pi {\rm d}} \nonumber \\
\times \left[ {\rm r}^0_{\varphi_0} 
e^{p(2\varphi_0-\varphi-\varphi_L)} + e^{p(\varphi-\varphi_L )} \right], 
\end{eqnarray}
\begin{eqnarray} 
\zeta^n_0(y')= \int_0^{\pi} d \varphi (1-{\rm r}^0_{\varphi}) 
\frac{\sin \varphi (\cos \varphi'_0)^n}{{\rm d}_0 \sin \varphi'_0} \nonumber \\
\times \left[ e^{p(\varphi-\varphi'_0)} +  e^{p(\varphi'_0-\varphi)} 
(1-{\rm d}_0)/{\rm r}^0_{\varphi} \right],
\end{eqnarray}
\begin{eqnarray} 
\zeta^n_L(y')= \int_0^{\pi} d \varphi (1-{\rm r}^L_{\varphi}) 
\frac{\sin \varphi (\cos \varphi'_L)^n }{{\rm d}_L \sin \varphi'_L} \nonumber \\
\times \left[ e^{p(\varphi'_L-\varphi)} +  e^{p(\varphi-\varphi'_L)} 
(1-{\rm d}_L)/{\rm r}^L_{\varphi} \right],
\end{eqnarray}
The functions $\zeta^n_{0}$ and $\zeta^n_{L}$ depend on $y'$ through $\varphi'_0$ and $\varphi'_L$. 

A transition to the limit $B=0$, when $R \rightarrow \infty$, is carried out by using the 
approximate expressions
\begin{eqnarray} 
\varphi-\varphi'=(y-y')/R \sin \varphi, \nonumber \\ 
\varphi_0=\varphi-y/R \sin \varphi,  \\
\varphi_L=\varphi-(y-L)/R \sin \varphi, \nonumber
\end{eqnarray}
valid at $R \gg |y-y'|$ (which always takes place at $R \gg L$) provided that 
$\sin \varphi$ is not very small. In this limit, the difference between 
$\varphi_0$, $\varphi_L$, $\varphi$, and $\varphi'$ goes to zero, but this 
difference still has to be taken into account in the exponential factors in order 
to compensate the large parameter $p=R/l$. Thus, in the lowest order in $L/R$, the 
quantity $R$ drops out from the equations and the general solution is written in the form  
\begin{eqnarray}
g_{\varphi}(y)= g_{\varphi}(0)e^{-y/l \sin \varphi} \nonumber \\
+ \frac{1}{\sin \varphi} \int_0^y dy' e^{(y'-y)/l \sin \varphi} {\cal F}_{\varphi}(y'), \\
g_{2\pi-\varphi}(y)= g_{2\pi-\varphi}(L)e^{(y-L)/l \sin \varphi} \nonumber \\ 
+ \frac{1}{\sin \varphi} \int_y^L dy' e^{(y-y')/l \sin \varphi} {\cal F}_{\varphi}(y'),
\end{eqnarray}
with $\varphi \in [0,\pi]$. The boundary conditions are simplified at $B=0$, since the 
integral terms $M_0$ and $M_L$ disappear. From these conditions, we determine the 
constants $g_{\varphi}(0)$ and $g_{2\pi-\varphi}(L)$ and obtain
\begin{eqnarray}
g_{\varphi}(y)= \int_{0}^{L} d y' Q^0_{\varphi}(y,y') {\cal F}_{\varphi}(y'), \nonumber \\
g_{2\pi-\varphi}(y)= \int_{0}^{L} d y' Q^1_{\varphi}(y,y') {\cal F}_{\varphi}(y'),
\end{eqnarray}
where
\begin{eqnarray}
Q^0_{\varphi}(y,y')= \left\{
[ \theta(y-y')+(1-{\rm d})/{\rm d} ] e^{(y'-y)/ l \sin \varphi}  \right. \nonumber \\ 
\left.  + {\rm r}^0_{\varphi} e^{-(y+y')/ l \sin \varphi}/{\rm d} \right\} \frac{1}{\sin \varphi} , \\
Q^1_{\varphi}(y,y')= \left\{
[ \theta(y'-y)+(1-{\rm d})/{\rm d} ] e^{(y-y')/ l \sin \varphi}  \right. \nonumber \\ 
\left. + {\rm r}^L_{\varphi} e^{(y+y'-2L)/ l \sin \varphi}/{\rm d} \right\} \frac{1}{\sin \varphi}, 
\end{eqnarray}
with
\begin{eqnarray}
{\rm d}=1-{\rm r}^{0}_{\varphi} {\rm r}^{L}_{\varphi} \lambda^2_{\varphi}, 
~~~\lambda_{\varphi}=e^{-L/l \sin \varphi}.
\end{eqnarray}
The integral equation (15) is reduced to 
\begin{eqnarray}
g_0(y)= \frac{1}{l} \int_{0}^{L} d y' {\cal K}_{00}(y,y') g_0(y'), \nonumber
\end{eqnarray}
which has a trivial solution $g_0(y)=0$. Strictly speaking, an arbitrary constant also 
satisfies this equation, but this is not essential because $g_0(y)$ is defined with the accuracy 
up to a constant. The integral equation (16) is reduced to Eq. (17) for $g_1(y)$. There, 
\begin{eqnarray}
{\cal K}_{11}(y,y')= \int_0^{\pi} \frac{d \varphi}{\pi} \frac{\cos^2 \varphi}{\sin \varphi}
\left[ e^{-|y-y'|/ l \sin \varphi} \right. \nonumber \\
 + 2  \cosh(|y-y'|/ l \sin \varphi) (1-{\rm d})/{\rm d} \nonumber \\
\left. +  {\rm r}^0_{\varphi} 
e^{-(y+y')/ l \sin \varphi}/{\rm d} + {\rm r}^L_{\varphi} e^{(y+y'-2L)/ l \sin \varphi}/{\rm d} \right]
\end{eqnarray}
and 
\begin{eqnarray}
{\cal L}_{1}(y)= l \int_0^{\pi} \frac{d \varphi}{\pi}  
\cos^2 \varphi [ 2
- \xi^L_{\varphi} e^{(y-L)/ l \sin \varphi}  ~~~~~~\\
-\xi^0_{\varphi} e^{-y/ l \sin \varphi}],~
\xi^{0,L}_{\varphi}=[1-{\rm r}^{0,L}_{\varphi}(1-\lambda_{\varphi})-{\rm r}^{0}_{\varphi} 
{\rm r}^{L}_{\varphi} \lambda_{\varphi}]/{\rm d}, \nonumber
\end{eqnarray}
in accordance with \cite{dejong}.

The formalism given above also allows for treatment of a semi-infinite plane, 
when a single boundary at $y=0$ is present. In this case, one should formally put 
$\varphi_L=\pi$, ${\rm r}^L_{\varphi_L}=1$, and extend the upper limit of the 
integration over $y'$ to infinity. The factor $M_L$ does not enter the distribution 
function because it always stands at $1-{\rm r}^L_{\varphi_L}$, while $M_0$, in view of 
$\beta_0=\beta_L=0$, is given by $M_0=a_{00} \kappa_0$, $a_{00}=1/({\cal N}_0-\alpha_0)$. 
Notice that in this case
\begin{eqnarray} 
{\cal N}_0-\alpha_0= \int_0^{\pi} d \varphi (1-{\rm r}^0_{\varphi}) \sin \varphi 
(1-e^{2p(\varphi-\pi)})/{\rm d}_0, \nonumber \\
{\rm d}_0=1-{\rm r}^0_{\varphi}e^{2p(\varphi-\pi)}.~~~~~
\end{eqnarray} 
In Eqs. (A20) and (A21), only the parts containing $Q^+_{\varphi}(y,y')$ and $a_{00}$ survive. 

Far from the boundaries, the electrons are moving in the cyclotron orbits and do not feel the boundaries.
Formally, this case is described by the substitutions $\varphi_0=0$, $\varphi_L=\pi$, ${\rm r}^0_{\varphi_0}=1$, 
and ${\rm r}^L_{\varphi_L}=1$ so that   
\begin{eqnarray} 
{\cal K}_{00}(y,y') = \int_0^{\pi} \frac{d \varphi}{2 \pi} \frac{\Lambda_{\varphi \varphi'}}{
\sin \varphi'}, ~~ \Lambda_{\varphi \varphi'}=\frac{1}{{\rm d}}[ e^{-p|\varphi'-\varphi|} \nonumber \\ 
+e^{p(|\varphi'-\varphi|-2 \pi)} + e^{-p(\varphi'+\varphi)}+
e^{p(\varphi'+\varphi -2 \pi)}], 
\end{eqnarray}
where ${\rm d}=1-e^{-2 \pi p}$ is constant. The other kernels, ${\cal K}_{10}(y,y')$, 
${\cal K}_{01}(y,y')$, and ${\cal K}_{11}(y,y')$ are given by the same expression with 
extra multipliers $2 \cos \varphi$, $\cos \varphi'$, and $2 \cos \varphi \cos \varphi'$ 
under the integral, respectively. As already mentioned, ${\cal K}_{nn'}(y,y')$ are nonzero 
only at $|y'-y|< 2R$. Using the identity   
\begin{eqnarray} 
\int_0^{\pi} d \varphi \Lambda_{\varphi \varphi'} = \int_0^{\pi} d \varphi' 
\Lambda_{\varphi \varphi'} = \frac{2}{p}, 
\end{eqnarray}
one can show that Eqs. (15) and (16) are satisfied for linear Hall voltage, $g_0(y)=C + eE y l_1/R$, 
and constant current, $g_1(y) =eE l_1$. This means that in the bulk of the sample the current and the Hall 
field are the same as in an infinitely wide sample, while within the layers of widths $4R$ near the boundaries 
the current and the Hall field are coordinate-dependent. 

\end{appendix}

\end{document}